\documentclass{article}

\usepackage[utf8]{inputenc}
\usepackage[T1]{fontenc}
\usepackage[margin=1in]{geometry}
\usepackage[numbers]{natbib}
\usepackage{url}
\usepackage{booktabs}
\usepackage{array}
\usepackage{tabularx}
\usepackage{amsmath}
\usepackage{microtype}
\usepackage{xcolor}
\usepackage{graphicx}
\usepackage{float}
\usepackage{placeins}
\usepackage{hyperref}

\hypersetup{colorlinks=true, linkcolor=blue, citecolor=blue, urlcolor=blue}
\newcolumntype{Y}{>{\raggedright\arraybackslash}X}
\graphicspath{{figures/}}

\title{Choosing a JPEG Decoder for PyTorch DataLoaders:\\
Workload-Specific Throughput on Four CPUs}

\author{Vladimir Iglovikov\\
Ternaus\\
\texttt{vladimir@ternaus.com}
\and
Dmitry Kosarevsky\\
Independent Researcher\\
\texttt{if.kosarevsky@gmail.com}}

\begin{document}

\maketitle

\begin{abstract}
A JPEG decoder benchmark can combine worker counts, CPUs, and datasets in one
large result matrix.  We simplify that comparison by fixing a PyTorch
\texttt{DataLoader} at eight workers and asking one question: how much faster is
each decoder than Pillow on the same CPU and JPEG workload?

We benchmark 12 Python decoders on four 16-vCPU Google Cloud
platforms and two workloads from the Forchheim Image Database.  The first
contains 324 large camera originals.  The second contains 1,668 originals and
copies processed by social media services.  Every decoder--CPU--workload setting has five
isolated repetitions.  JPEG bytes reside in memory, and each timed decode
returns a complete three-channel RGB array.

The workload changes the practical result.  On camera originals, the geometric
mean across CPUs ranges from \(1.00\times\) to \(1.09\times\) Pillow.  On the
mixed workload, \texttt{simplejpeg}, \texttt{imagecodecs}, and
\texttt{turbojpeg} reach \(1.39\)--\(1.40\times\) Pillow.  The exact leader
still changes by CPU.  We therefore recommend comparing decoders on
representative JPEGs, while using the fixed eight-worker results as a compact
starting point.  Full worker curves and absolute throughput tables are in the
appendix.
\end{abstract}

\section{Introduction}

The JPEG decoder is one part of a training input pipeline.  If that pipeline
delivers the next batch too late, the GPU waits.  A faster decoder can help in
that case.  The useful engineering question is how much decoder choice changes
the rate at which a \texttt{DataLoader} supplies images.

This question is easy to bury in a benchmark matrix.  Throughput changes with
the decoder, worker count, CPU, and JPEG distribution.  Reporting every
combination gives a complete record, but makes the main result hard to see.
We reduce the main comparison in three steps:
\begin{enumerate}
  \item We discuss the two JPEG workloads separately.
  \item We fix \texttt{num\_workers} at eight for every decoder and CPU.
  \item We divide each result by Pillow throughput on the same CPU and
  workload, then summarize the four CPU-specific ratios with a geometric mean.
\end{enumerate}
Absolute images per second remain available in Appendix~\ref{sec:absolute}.
The fixed worker count is a comparison convention.  It does not claim that
eight workers minimizes training time.

The two workloads separate clearly.  Decoder choice has a modest effect on the
large camera originals and a much larger effect on the mixture of originals
and smaller social-media copies.  Hardware changes the exact ranking, but it
does not erase this workload difference.

\paragraph{Contributions.}
\begin{enumerate}
  \item We compare 12 CPU JPEG decoders at one shared DataLoader setting on
  four 16-vCPU platforms and two real JPEG workloads.
  \item We divide each result by Pillow throughput on the same CPU, so hardware
  speed does not dominate the decoder comparison.
  \item We publish the full 3,725-bundle evidence archive, including worker
  sweeps, package versions, exact plans, and validation code.
\end{enumerate}

\section{Related work}

FFCV~\citep{leclerc2023ffcv}, WebDataset~\citep{benezitan2021webdataset},
TorchData~\citep{torchdata2024}, NVIDIA DALI~\citep{nvidia2024dali}, and
nvJPEG~\citep{nvidia2024nvjpeg} address storage layout, sharding,
preprocessing, or accelerator-side decoding.  We study a narrower path: JPEG
bytes decoded on the CPU and delivered through a PyTorch-style loader.  The
JPEG files are already in memory when timing starts, so storage speed does not
enter the comparison.

Python pipelines expose several JPEG implementations through Pillow
\citep{pillow2024}, OpenCV~\citep{opencv2024}, torchvision
\citep{pytorch2024,torchvision2024}, Kornia~\citep{kornia2024}, and wrappers
around libjpeg-turbo~\citep{libjpeg-turbo}.  Their backends, bindings,
memory allocation, and thread controls differ.  Our adapters convert every
successful result to the same array format before counting the image as
decoded.

\section{Experimental setup}

\subsection{Two FODB workloads}

The Forchheim Image Database (FODB) contains camera originals and copies
processed by social-media services~\citep{hadwiger2021fodb}.  We select 12
scenes from all 27 cameras with a seeded ordering.

\textbf{FODB-native} contains 324 camera originals.  The median image has
12.19 megapixels and occupies 3.74 MiB as a JPEG.  This workload represents
large, camera-generated files.

\textbf{FODB-mixed} starts from the same camera--scene pairs and adds copies
from Facebook, Instagram, Telegram, Twitter, and WhatsApp.  The timed common
set contains 1,668 images with a median of 1.23 megapixels and 0.24 MiB.  The
mixture changes resolution, compression, metadata, and encoding history
together.  It supports a workload comparison, not a causal claim about any one
JPEG property.

\subsection{Decoders, CPUs, and output}

We test \texttt{ajpegli}, \texttt{imagecodecs}, \texttt{imageio},
\texttt{jpeg4py}, \texttt{kornia-rs}, OpenCV, Pillow, PyVips, scikit-image,
\texttt{simplejpeg}, \texttt{torchvision}, and \texttt{turbojpeg}.  OpenCV,
PyVips, and \texttt{torchvision} are explicitly limited to one internal thread.
Runtime checks report one effective thread for the other decoders.

The four Google Cloud platforms each provide 16 vCPUs.  The three x86
platforms are \texttt{c4-standard-16} with Intel Xeon 8581C,
\texttt{c3d-standard-16} with AMD Zen~4, and
\texttt{c4d-standard-16} with AMD Zen~5.  The Arm platform is
\texttt{c4a-standard-16} with Google Axion / Neoverse V2
\citep{gcp2026c4}.  Package versions are identical across platforms.

Every successful decode returns a complete C-contiguous RGB
\texttt{uint8} array with shape \texttt{(height, width, 3)}.  Compressed JPEG
bytes are loaded into memory before validation, warmup, or timing.  The loader
uses batch size 1, pass-through collation, the \texttt{spawn} start method,
prefetch factor 1, and persistent workers.

We assume that users have already checked that their chosen decoder accepts
their JPEGs and returns the color format their application expects.\footnote{A
small compatibility audit is reported in Appendix~\ref{sec:compatibility}.  It
covers one malformed JPEG pattern and one four-component JPEG.  These cases
cannot establish general decoder reliability.}  The main results compare speed
on the same input images for every decoder.

\subsection{Fixed comparison and summary statistic}

The main analysis uses eight DataLoader workers for every decoder, CPU, and
workload.  This gives 480 timed bundles:
\[
  12\ \text{decoders} \times 4\ \text{CPUs} \times
  2\ \text{workloads} \times 5\ \text{repetitions}.
\]
Each repetition runs in a fresh process and times one complete traversal.

For decoder \(d\), CPU \(c\), workload \(s\), and repetition \(r\), we report
the paired ratio
\[
  R_{d,c,s,r} =
  \frac{\text{images/s}_{d,c,s,r}}
       {\text{images/s}_{\mathrm{Pillow},c,s,r}}.
\]
The figures show all five paired ratios on each CPU.  A black diamond marks the
geometric mean of the four CPU-specific mean ratios.  This normalization keeps
hardware speed from dominating the decoder comparison.  It does not mean that
the four CPUs behave the same.

\section{Camera originals: decoder choice changes throughput modestly}
\label{sec:native}

The decoder gaps are small on camera originals.  Across the four CPUs, the
geometric mean ranges from \(1.00\times\) for Pillow to \(1.09\times\) for
\texttt{torchvision}.  PyVips follows at \(1.08\times\).  The other nine
decoders fall between \(1.02\times\) and \(1.06\times\).

\begin{figure}[H]
  \centering
  \includegraphics[width=\linewidth]{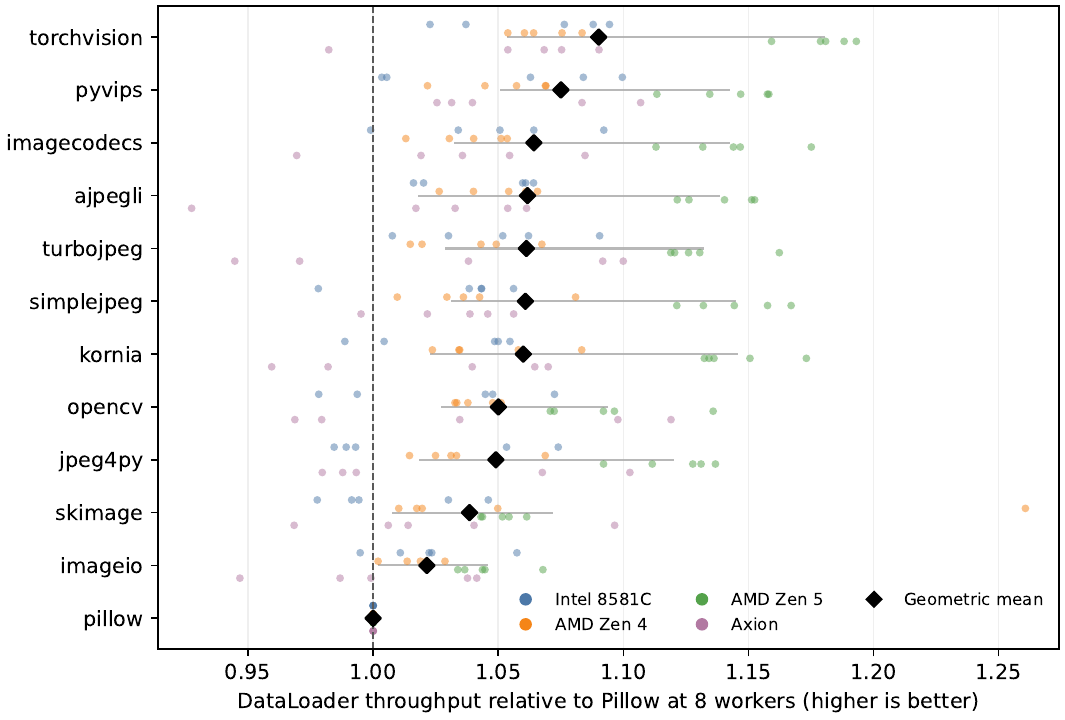}
  \caption{DataLoader throughput on 324 FODB camera originals at eight workers,
  relative to Pillow on the same 16-vCPU platform and repetition.  Colored
  points show five paired repetitions on each CPU.  Gray lines span the four
  CPU-specific mean ratios; black diamonds show their geometric mean.  Every
  timed output is a complete RGB \texttt{uint8} array.}
  \label{fig:native-w8}
\end{figure}

The exact leader depends on the CPU.  The fastest mean is
\texttt{torchvision} on Intel and Zen~5, scikit-image on Zen~4, and PyVips on
Axion.  The top mean exceeds Pillow by about 6\%--7\% on Intel, Zen~4, and
Axion, and by 18\% on Zen~5.  Absolute leader throughput ranges from 17.4 to
40.1 images/s across the four machines.

For this workload, decoder selection offers a limited gain compared with the
variation between CPUs.  A user processing similar camera originals can keep
Pillow as a reasonable baseline and benchmark a small shortlist only when
input throughput is already a measured bottleneck.

\FloatBarrier

\section{Mixed JPEGs: the leading decoders reach about 1.4 times Pillow}
\label{sec:mixed}

\begin{figure}[H]
  \centering
  \includegraphics[width=\linewidth]{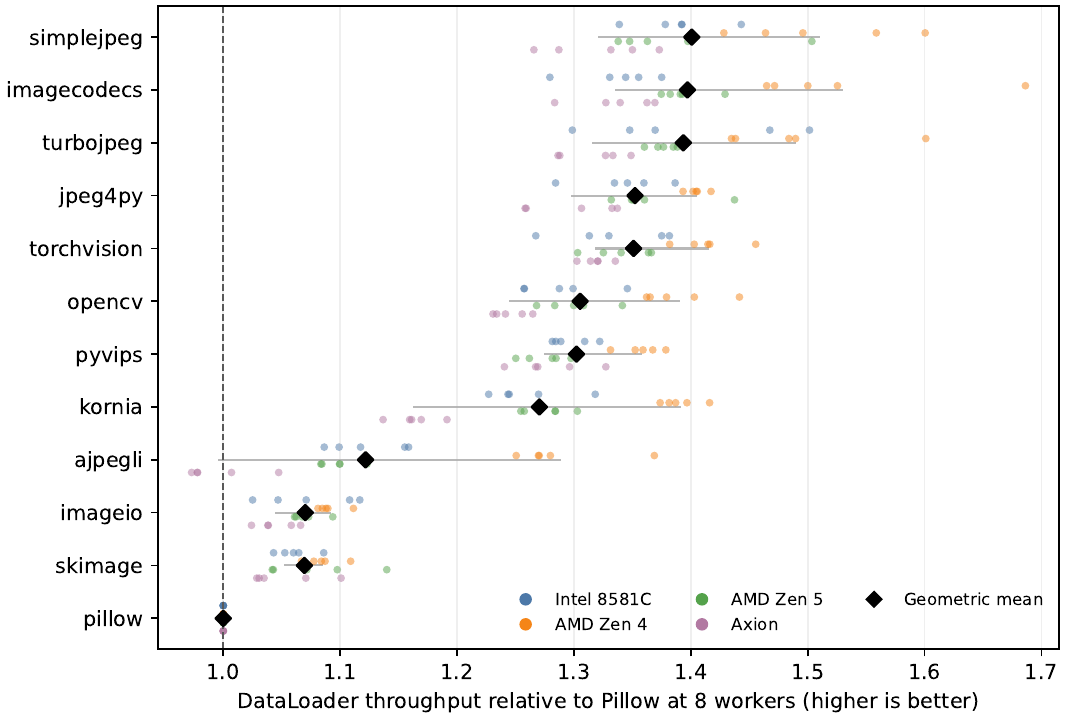}
  \caption{DataLoader throughput on 1,668 FODB originals and copies processed
  by social media services at eight workers, relative to Pillow on the
  same 16-vCPU platform and repetition.  Colored points show five paired
  repetitions on each CPU.  Gray lines span the four CPU-specific mean ratios;
  black diamonds show their geometric mean.}
  \label{fig:mixed-w8}
\end{figure}

The mixed workload produces a clearer separation.  The top group in
Figure~\ref{fig:mixed-w8} contains \texttt{simplejpeg}, \texttt{imagecodecs},
and \texttt{turbojpeg}.  Their geometric means are \(1.40\times\), \(1.40\times\), and
\(1.39\times\) Pillow.  Their CPU-specific mean ratios span
\(1.32\times\)--\(1.53\times\).

The leading library still changes by CPU.  \texttt{turbojpeg} has the highest
mean on Intel, while \texttt{imagecodecs} leads on Zen~4, Zen~5, and Axion.
The gap within the top three is small compared with their gap over Pillow.
Their absolute means range from about 58 to 110 images/s, depending on CPU.

The mixed workload changes several JPEG properties at once, so this experiment
cannot identify the cause of the larger gap.  The measured gap still shows
that a decoder choice which matters little for camera originals can matter much
more for another real JPEG population.

\FloatBarrier

\section{How to use these results}

Start by matching the workload.  For large camera originals, decoder choice
changed the four-CPU geometric mean by at most 9\% relative to Pillow.  For the
mixed workload, test \texttt{simplejpeg}, \texttt{imagecodecs}, and
\texttt{turbojpeg} first.  Keep Pillow in the comparison so the ratio remains
easy to interpret.

Then run the shortlist on the deployment CPU and representative JPEGs.
Hardware changes the exact leader, and a new JPEG distribution may change the
ranking again.

Eight workers make the decoder comparison compact.  They are a starting point,
not a tuned training setting.  If the model still waits for input, test nearby
worker counts in the real pipeline and measure GPU idle time and wall-clock
training time.  Appendix~\ref{sec:workers} reports our wider worker sweep and
shows why its result should not be treated as a universal default.

\section{Limitations}

FODB-native and FODB-mixed differ in resolution, compressed size, estimated
quality, progressive encoding, source service, and item count.  The experiment
compares these two workloads as wholes.  It cannot assign their different
decoder gaps to one property.

The common-input rule removes 276 WhatsApp files from the mixed timing set
because not every decoder accepts them.  The timed population therefore
contains 1,668 of the planned 1,944 images.  This preserves a fair speed
comparison, but changes the WhatsApp share.

The fixed eight-worker result measures DataLoader throughput, not training speed.
The benchmark excludes storage, augmentation, tensor stacking, pinned-memory
transfer, GPU decode, and model computation.  Batch size 1 avoids padding
images of different sizes and is not a proposed training default.

The hardware scope is four 16-vCPU Google Cloud machine types running the same
Linux software environment.  Five repetitions show run-to-run variation but
cannot resolve small differences between nearly equal means.

\section{Reproducibility}

The public artifact is available at
\url{https://github.com/ternaus/imread_benchmark}.  It contains content-addressed
datasets, benchmark plans, run bundles, validation code, and generated figures
and tables.  The full archive has 3,725 validated FODB speed bundles.  The
fixed eight-worker result in this paper is a balanced 480-bundle subset.
Appendix~\ref{sec:evidence-receipt} records the package and plan identities.

\section{Conclusion}

JPEG decoder choice depends on the workload.  At eight DataLoader workers, the
12 tested decoders stay within 9\% of Pillow by four-CPU geometric mean on large
camera originals.  On the mixed workload, the top three reach about
\(1.4\times\) Pillow.  Normalize within each CPU, inspect each workload
separately, and use the result to choose a small local benchmark shortlist.

\section*{Acknowledgments}

We thank Google and the Google Developer Program for providing Google Cloud
credits that supported the benchmark runs.

\bibliographystyle{plainnat}
\bibliography{references_preprint}

@inproceedings{hadwiger2021fodb,
  author    = {Hadwiger, Benjamin and Riess, Christian},
  title     = {The Forchheim Image Database for Camera Identification in the Wild},
  booktitle = {Pattern Recognition. ICPR International Workshops and Challenges},
  series    = {Lecture Notes in Computer Science},
  pages     = {500--515},
  publisher = {Springer},
  year      = {2021},
  doi       = {10.1007/978-3-030-68780-9_40}
}

@inproceedings{leclerc2023ffcv,
  title     = {{FFCV}: Accelerating Training by Removing Data Bottlenecks},
  author    = {Leclerc, Guillaume and Ilyas, Andrew and Engstrom, Logan and Park, Sung Min and Salman, Hadi and M{\k{a}}dry, Aleksander},
  booktitle = {Proceedings of the IEEE/CVF Conference on Computer Vision and Pattern Recognition (CVPR)},
  pages     = {12011--12020},
  year      = {2023}
}

@misc{benezitan2021webdataset,
  title        = {WebDataset: a {PyTorch} Dataset ({WebDataset}) designed for streaming training},
  author       = {Bob McElrath and Thomas Breuel},
  year         = {2021},
  howpublished = {\url{https://github.com/webdataset/webdataset}},
  note         = {Accessed 2026-05-02}
}

@misc{nvidia2024dali,
  title        = {{NVIDIA DALI}: {GPU}-accelerated data loading and image augmentation},
  author       = {{NVIDIA Corporation}},
  year         = {2024},
  howpublished = {\url{https://developer.nvidia.com/dali}},
  note         = {Accessed 2026-05-02}
}

@misc{nvidia2024nvjpeg,
  title        = {{nvJPEG}: {GPU}-accelerated {JPEG} decode},
  author       = {{NVIDIA Corporation}},
  year         = {2024},
  howpublished = {\url{https://developer.nvidia.com/nvjpeg}},
  note         = {Accessed 2026-05-02}
}

@misc{torchdata2024,
  title        = {TorchData},
  author       = {{PyTorch Team}},
  year         = {2024},
  howpublished = {\url{https://github.com/pytorch/data}},
  note         = {Accessed 2026-05-02}
}

@misc{libjpeg-turbo,
  title        = {libjpeg-turbo},
  author       = {{The libjpeg-turbo Project}},
  year         = {2024},
  howpublished = {\url{https://libjpeg-turbo.org}},
  note         = {Accessed 2026-05-02}
}

@misc{pillow2024,
  title        = {{Pillow}: the friendly {PIL} fork},
  author       = {{Pillow Developers}},
  year         = {2024},
  howpublished = {\url{https://pillow.readthedocs.io/en/stable/}},
  note         = {Accessed 2026-05-02}
}

@misc{opencv2024,
  title        = {Open Source Computer Vision Library ({OpenCV})},
  author       = {{OpenCV Team}},
  year         = {2024},
  howpublished = {\url{https://opencv.org}},
  note         = {Accessed 2026-05-02}
}

@misc{pytorch2024,
  title        = {{PyTorch}},
  author       = {{PyTorch Team}},
  year         = {2024},
  howpublished = {\url{https://pytorch.org}},
  note         = {Accessed 2026-05-02}
}

@misc{torchvision2024,
  title        = {torchvision},
  author       = {{PyTorch Team}},
  year         = {2024},
  howpublished = {\url{https://pytorch.org/vision}},
  note         = {Accessed 2026-05-02}
}

@misc{kornia2024,
  title        = {Kornia: differentiable computer vision in {PyTorch}},
  author       = {Riba, Edgar and Mishkin, Dmytro and Ponsa, Daniel and Rublee, Ethan and Bradski, Gary},
  year         = {2024},
  howpublished = {\url{https://kornia.github.io}},
  note         = {Accessed 2026-05-02}
}

@misc{gcp2026c4,
  title        = {{Google Cloud} Compute Engine machine families (C4 / C4D / C4A / T2A documentation)},
  author       = {{Google Cloud}},
  year         = {2026},
  howpublished = {\url{https://cloud.google.com/compute/docs/machine-types}},
  note         = {Accessed 2026-05-02}
}

\appendix
\section{Absolute throughput at eight workers}
\label{sec:absolute}

Tables~\ref{tab:native-w8} and~\ref{tab:mixed-w8} give the absolute values
behind Figures~\ref{fig:native-w8} and~\ref{fig:mixed-w8}.  Each cell is the
mean and sample standard deviation over five isolated repetitions.  The final
column is the geometric mean of the four CPU-specific paired ratios to Pillow.

\begin{table}[H]
  \centering
  \small
  \setlength{\tabcolsep}{4pt}
  \renewcommand{\arraystretch}{1.16}
  \begin{tabular}{lrrrrr}
    \textbf{Decoder} & \textbf{Intel} & \textbf{Zen 4} & \textbf{Zen 5} &
    \textbf{Axion} & \textbf{Geo. mean} \\
    \hline
    \texttt{torchvision} & $18.7 \pm 0.6$ & $17.4 \pm 0.1$ & $40.1 \pm 0.5$ & $28.2 \pm 0.7$ & 1.09$\times$ \\
\texttt{pyvips} & $18.4 \pm 0.9$ & $17.1 \pm 0.3$ & $38.8 \pm 0.4$ & $28.3 \pm 0.4$ & 1.08$\times$ \\
\texttt{imagecodecs} & $18.4 \pm 0.5$ & $16.9 \pm 0.3$ & $38.8 \pm 0.5$ & $27.6 \pm 0.2$ & 1.06$\times$ \\
\texttt{ajpegli} & $18.3 \pm 0.6$ & $17.1 \pm 0.2$ & $38.7 \pm 0.6$ & $27.2 \pm 0.9$ & 1.06$\times$ \\
\texttt{turbojpeg} & $18.4 \pm 0.7$ & $16.9 \pm 0.2$ & $38.4 \pm 0.5$ & $27.5 \pm 1.1$ & 1.06$\times$ \\
\texttt{simplejpeg} & $18.1 \pm 0.6$ & $16.9 \pm 0.3$ & $38.9 \pm 0.5$ & $27.6 \pm 0.9$ & 1.06$\times$ \\
\texttt{kornia} & $18.0 \pm 0.6$ & $17.0 \pm 0.3$ & $38.9 \pm 0.5$ & $27.4 \pm 0.7$ & 1.06$\times$ \\
\texttt{opencv} & $18.0 \pm 0.7$ & $17.0 \pm 0.1$ & $37.1 \pm 0.7$ & $27.8 \pm 1.0$ & 1.05$\times$ \\
\texttt{jpeg4py} & $17.9 \pm 0.8$ & $16.9 \pm 0.3$ & $38.0 \pm 0.4$ & $27.4 \pm 0.8$ & 1.05$\times$ \\
\texttt{skimage} & $17.7 \pm 0.5$ & $17.4 \pm 1.7$ & $35.7 \pm 0.3$ & $27.4 \pm 0.8$ & 1.04$\times$ \\
\texttt{imageio} & $17.9 \pm 0.3$ & $16.6 \pm 0.3$ & $35.5 \pm 0.4$ & $26.8 \pm 0.8$ & 1.02$\times$ \\
\texttt{pillow} & $17.5 \pm 0.3$ & $16.3 \pm 0.2$ & $34.0 \pm 0.3$ & $26.8 \pm 1.0$ & 1.00$\times$ 

  \end{tabular}
  \caption{FODB-native DataLoader throughput in images/s at eight workers.
  Higher is better.}
  \label{tab:native-w8}
\end{table}

\begin{table}[H]
  \centering
  \small
  \setlength{\tabcolsep}{4pt}
  \renewcommand{\arraystretch}{1.16}
  \begin{tabular}{lrrrrr}
    \textbf{Decoder} & \textbf{Intel} & \textbf{Zen 4} & \textbf{Zen 5} &
    \textbf{Axion} & \textbf{Geo. mean} \\
    \hline
    \texttt{simplejpeg} & $59.6 \pm 2.4$ & $59.1 \pm 2.7$ & $110.1 \pm 4.3$ & $87.2 \pm 3.0$ & 1.40$\times$ \\
\texttt{imagecodecs} & $57.3 \pm 1.7$ & $59.9 \pm 3.6$ & $110.5 \pm 2.8$ & $88.2 \pm 2.3$ & 1.40$\times$ \\
\texttt{turbojpeg} & $59.8 \pm 2.6$ & $58.3 \pm 2.8$ & $109.1 \pm 2.0$ & $86.8 \pm 1.8$ & 1.39$\times$ \\
\texttt{jpeg4py} & $57.5 \pm 1.5$ & $55.0 \pm 0.6$ & $108.2 \pm 2.3$ & $85.6 \pm 1.4$ & 1.35$\times$ \\
\texttt{torchvision} & $57.2 \pm 2.0$ & $55.4 \pm 1.2$ & $106.2 \pm 0.9$ & $87.0 \pm 1.2$ & 1.35$\times$ \\
\texttt{opencv} & $55.3 \pm 1.1$ & $54.4 \pm 1.7$ & $103.0 \pm 2.5$ & $82.1 \pm 0.8$ & 1.31$\times$ \\
\texttt{pyvips} & $55.6 \pm 1.3$ & $53.1 \pm 0.3$ & $101.1 \pm 1.3$ & $84.4 \pm 2.8$ & 1.30$\times$ \\
\texttt{kornia} & $54.0 \pm 1.0$ & $54.4 \pm 0.7$ & $101.2 \pm 1.2$ & $76.7 \pm 1.0$ & 1.27$\times$ \\
\texttt{ajpegli} & $48.2 \pm 1.7$ & $50.4 \pm 1.7$ & $87.0 \pm 1.9$ & $65.7 \pm 0.9$ & 1.12$\times$ \\
\texttt{imageio} & $46.0 \pm 1.7$ & $42.7 \pm 0.6$ & $84.9 \pm 1.9$ & $68.9 \pm 0.7$ & 1.07$\times$ \\
\texttt{skimage} & $45.5 \pm 1.0$ & $42.5 \pm 0.5$ & $85.5 \pm 3.5$ & $69.4 \pm 1.2$ & 1.07$\times$ \\
\texttt{pillow} & $42.9 \pm 1.0$ & $39.1 \pm 0.4$ & $79.3 \pm 1.2$ & $66.0 \pm 1.3$ & 1.00$\times$ 

  \end{tabular}
  \caption{FODB-mixed DataLoader throughput in images/s at eight workers.
  Higher is better.}
  \label{tab:mixed-w8}
\end{table}

\section{Worker-count sensitivity}
\label{sec:workers}

The full experiment starts with 0, 2, 4, and 8 workers.  A rule fixed before
the next stage adds 12 workers when the highest base-grid mean occurs at eight.
This selects 87 of 96 decoder--CPU--workload combinations.  A second rule adds
16 workers when the 12-worker mean is at least 5\% above the 8-worker mean and
all five paired repetitions avoid a decrease.  This selects 58 combinations.

The mixed workload reaches its highest observed mean at 16 workers in 43 of 48
decoder--CPU combinations and at 12 in the other five.  Native images peak at
12 workers in 35 combinations, zero in nine, eight in two, and 16 in two.
These are DataLoader-only results.  Training workers share CPU and memory
with the rest of the pipeline, so the observed peaks do not define a training
default.

\begin{figure}[H]
  \centering
  \includegraphics[width=\linewidth]{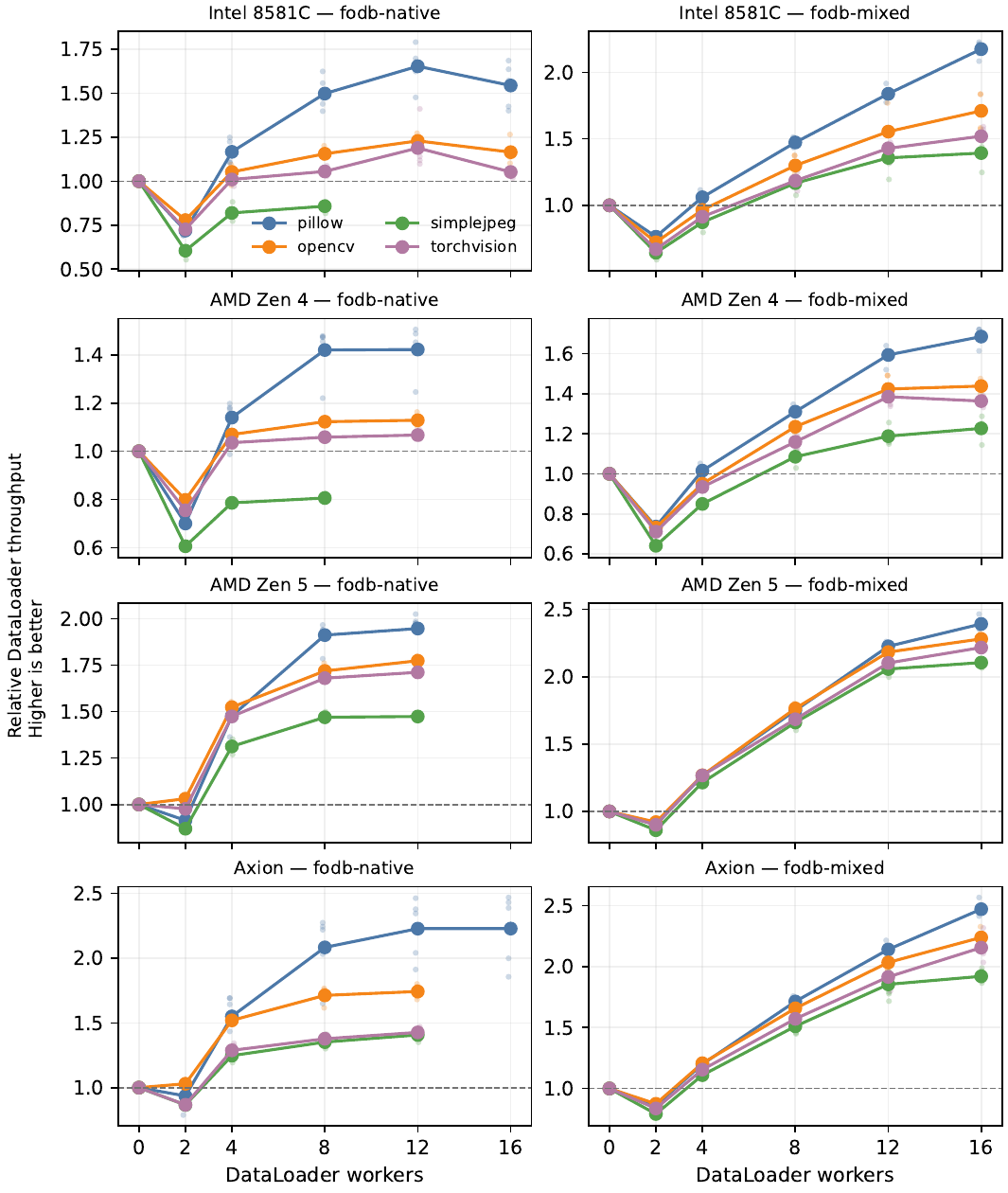}
  \caption{Worker scaling for four representative controlled-thread decoders.
  Throughput is divided by the same decoder at zero workers; higher is better.
  Lines are means of five repetition-paired ratios, and translucent points show
  each ratio.  Every curve contains 0, 2, 4, and 8 workers.  The staged rules
  determine whether 12- and 16-worker points appear.}
  \label{fig:worker-scaling}
\end{figure}

\begin{table}[H]
  \centering
  \small
  \setlength{\tabcolsep}{4pt}
  \begin{tabularx}{\linewidth}{lYYr}
    \textbf{Platform} & \textbf{Native} & \textbf{Mixed} & \textbf{Changed} \\
    \hline
    Intel 8581C & 0: 4, 12: 7, 16: 1 & 12: 1, 16: 11 & 11/12 \\
AMD Zen 4 & 0: 5, 8: 1, 12: 6 & 12: 3, 16: 9 & 11/12 \\
AMD Zen 5 & 8: 1, 12: 11 & 12: 1, 16: 11 & 12/12 \\
Axion & 12: 11, 16: 1 & 16: 12 & 11/12 

  \end{tabularx}
  \caption{Highest observed mean worker counts for the 12 controlled-thread
  decoders.  An entry such as 16: 11 means that 11 decoders peak at 16
  workers.  Changed counts decoders whose peak differs between workloads.}
  \label{tab:worker-transfer}
\end{table}

\section{One-thread choice versus a tuned DataLoader}
\label{sec:one-thread}

The fastest controlled one-thread decoder and the fastest decoder after worker
tuning differ in five of eight CPU--workload pairs.  Choosing the one-thread
leader and then tuning only its workers loses 0.9\%--7.9\% of mean DataLoader
throughput in those five pairs.  The losses stay within 10\%, but the complete
decoder ranking can change by a large margin.

\begin{table}[H]
  \centering
  \small
  \setlength{\tabcolsep}{4pt}
  \begin{tabular}{llllrr}
    \textbf{Workload} & \textbf{Platform} & \textbf{One-thread leader} &
    \textbf{Tuned leader} & \textbf{Loss} & \textbf{Rank corr.} \\
    \hline
    FODB-native & Intel 8581C & \texttt{turbojpeg} & \texttt{turbojpeg} (0 workers) & 0.0\% & 0.66 \\
FODB-native & AMD Zen 4 & \texttt{imagecodecs} & \texttt{imagecodecs} (0 workers) & 0.0\% & 0.90 \\
FODB-native & AMD Zen 5 & \texttt{imagecodecs} & \texttt{torchvision} (12 workers) & 5.0\% & 0.13 \\
FODB-native & Axion & \texttt{torchvision} & \texttt{jpeg4py} (12 workers) & 7.9\% & 0.23 \\
FODB-mixed & Intel 8581C & \texttt{imagecodecs} & \texttt{torchvision} (16 workers) & 0.9\% & 0.54 \\
FODB-mixed & AMD Zen 4 & \texttt{imagecodecs} & \texttt{imagecodecs} (12 workers) & 0.0\% & 0.77 \\
FODB-mixed & AMD Zen 5 & \texttt{imagecodecs} & \texttt{turbojpeg} (16 workers) & 3.6\% & 0.88 \\
FODB-mixed & Axion & \texttt{imagecodecs} & \texttt{torchvision} (16 workers) & 5.7\% & 0.75 

  \end{tabular}
  \caption{Fastest controlled one-thread decoder versus fastest tuned
  DataLoader decoder.  Lower loss is better.  Rank correlation covers all 12
  decoders.}
  \label{tab:protocol-decisions}
\end{table}

\begin{figure}[H]
  \centering
  \includegraphics[width=\linewidth]{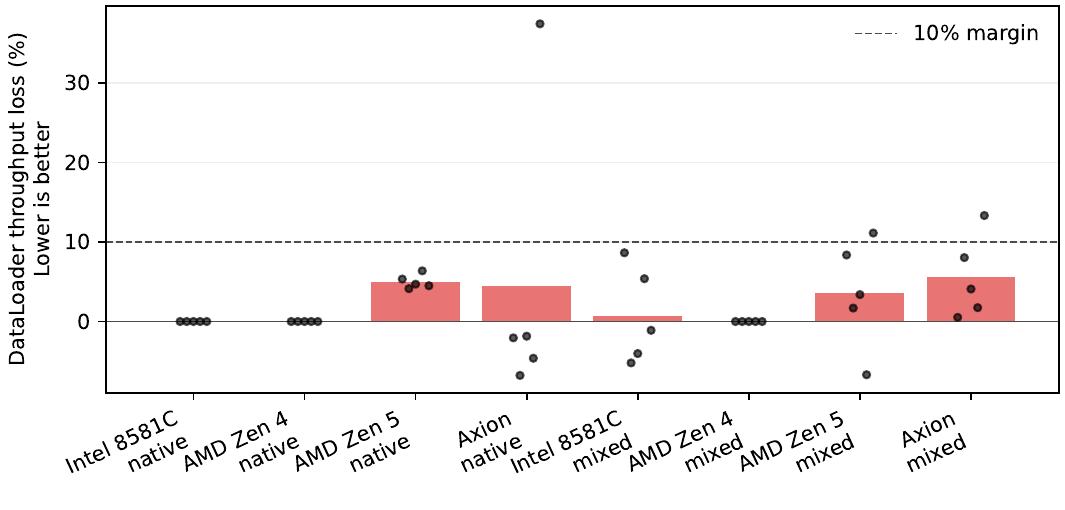}
  \caption{DataLoader throughput lost by choosing the fastest controlled
  one-thread decoder and then tuning its worker count.  Bars are mean losses;
  black points show five repetition-matched values.  Red marks a change in the
  fastest decoder.  The 10\% line is a practical margin, not a confidence
  threshold.}
  \label{fig:protocol-regret}
\end{figure}

\section{Compatibility scope}
\label{sec:compatibility}

The speed comparison assumes that the application has already validated its
JPEG population and required output.  We ran two narrow checks to document why
that assumption matters.  The first contains 276 progressive WhatsApp JPEGs
with empty DHT markers.  The second is one valid four-component JPEG and asks
whether the decoder returns the benchmark's three-channel RGB array.

\begin{table}[H]
  \centering
  \small
  \setlength{\tabcolsep}{8pt}
  \begin{tabular}{lrrr}
    \textbf{Decoder} & \textbf{Empty-DHT} & \textbf{Four-component} &
    \textbf{Both} \\
    \hline
    \texttt{ajpegli} & 0/276 & 0/1 & 0/277 \\
\texttt{imagecodecs} & 276/276 & 0/1 & 276/277 \\
\texttt{imageio} & 276/276 & 0/1 & 276/277 \\
\texttt{jpeg4py} & 276/276 & 0/1 & 276/277 \\
\texttt{kornia} & 276/276 & 0/1 & 276/277 \\
\texttt{opencv} & 276/276 & 1/1 & 277/277 \\
\texttt{pillow} & 276/276 & 1/1 & 277/277 \\
\texttt{pyvips} & 276/276 & 0/1 & 276/277 \\
\texttt{simplejpeg} & 276/276 & 1/1 & 277/277 \\
\texttt{skimage} & 276/276 & 0/1 & 276/277 \\
\texttt{torchvision} & 276/276 & 0/1 & 276/277 \\
\texttt{turbojpeg} & 276/276 & 0/1 & 276/277 

  \end{tabular}
  \caption{Results for two compatibility checks.  Entries show files passed
  out of files tested.  The checks cover two observed cases and do not measure
  general decoder reliability.}
  \label{tab:decoder-compatibility}
\end{table}

OpenCV, Pillow, and \texttt{simplejpeg} pass both checks under the pinned Linux
environment.  Other applications may need different color conversion, error
recovery, or fallback behavior.  They should repeat these checks on their own
data.  This small audit is not a general ranking.

\section{Timed workload composition}
\label{sec:workload-composition}

Every decoder is timed on the same images.  Native retains all 324 planned
files.  Mixed retains 1,668 of 1,944.  All 276 exclusions are progressive
WhatsApp copies with estimated quality 69.  The timed mixed set contains 324
images from the original, Facebook, Instagram, Telegram, and Twitter sources,
and 48 from WhatsApp.

\begin{figure}[H]
  \centering
  \includegraphics[width=\linewidth]{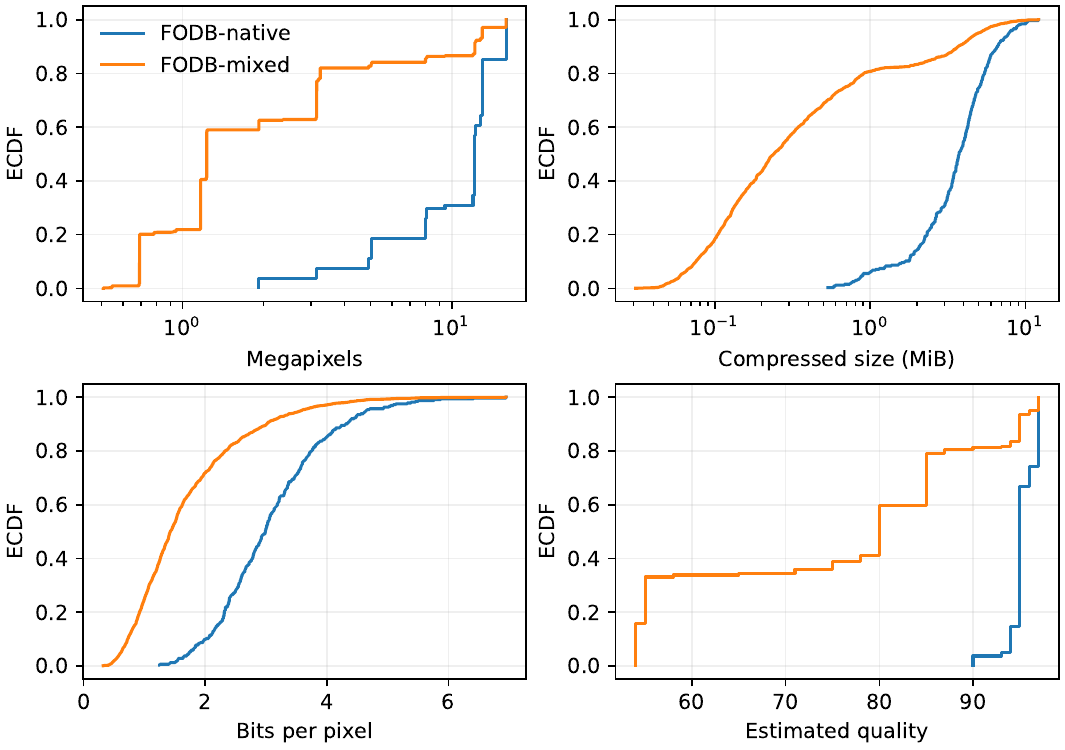}
  \caption{Empirical distributions for the two timed workloads.  Resolution,
  compressed size, bits per pixel, estimated quality, and progressive encoding
  change together.  The figure describes the workload shift and does not
  isolate a causal factor.}
  \label{fig:workload-distributions}
\end{figure}

\section{Evidence receipt and package versions}
\label{sec:evidence-receipt}

The complete speed archive contains 3,725 bundles: 1,745 FODB-native and 1,980
FODB-mixed.  Every selected configuration has five fresh-process repetitions.
The analysis validates the plan IDs, runner revision, package identities,
output contract, platform identity, repetition count, timed images, and
expected bundle count before generating paper assets.

\begin{description}
  \item[Dataset package] \nolinkurl{925d57ed84cb07e12f290e6c1612b61ca892fabea729091bfabd0acf3661cc9f}
  \item[Native base plan] \nolinkurl{6c746a5c21978f75c58c8c779f6628c2645a199b63489da9b9f1a9bb4f335c8f}
  \item[Mixed base plan] \nolinkurl{e04f54dc177cadf86cb99a9fc34b950afb5b1abee5b46af040bafc8a4aa7910b}
  \item[Runner revision] \nolinkurl{52a0bea5a2f44079883c9a41472b3add285955f2b471c37674ddd2fb5d4da6ff}
\end{description}

Table~\ref{tab:versions} lists the package versions recorded on every
platform.  The libjpeg-turbo build reports version 3.2.0, build 20260630.

\begin{table}[H]
  \centering
  \small
  \begin{tabular}{ll}
    \textbf{Distribution} & \textbf{Version} \\
    \hline
    \texttt{ajpegli} & \texttt{1.0.0} \\
\texttt{imagecodecs} & \texttt{2026.6.26} \\
\texttt{imageio} & \texttt{2.37.4} \\
\texttt{jpeg4py} & \texttt{0.1.4} \\
\texttt{kornia-rs} & \texttt{0.1.14} \\
\texttt{opencv-python-headless} & \texttt{5.0.0.93} \\
\texttt{pillow} & \texttt{12.3.0} \\
\texttt{pyturbojpeg} & \texttt{2.5.0} \\
\texttt{pyvips} & \texttt{3.1.1} \\
\texttt{scikit-image} & \texttt{0.26.0} \\
\texttt{simplejpeg} & \texttt{1.9.0} \\
\texttt{torch} & \texttt{2.13.0+cpu} \\
\texttt{torchvision} & \texttt{0.28.0+cpu} 

  \end{tabular}
  \caption{Pinned package versions for the FODB experiment.}
  \label{tab:versions}
\end{table}

\section{Thread-control check}

OpenCV, PyVips, and \texttt{torchvision} are measured both with package-default
thread behavior and with one requested internal thread.  Default PyVips reports
up to 16 effective threads in the one-thread decode protocol, and default
\texttt{torchvision} reports up to eight.  The main comparison therefore uses
the explicit one-thread configurations for these three libraries.  Default
thread results remain in the generated evidence for applications that preserve
package defaults.

\end{document}